\documentclass[twocolumn,aps,prd,showpacs,preprintnumbers]{revtex4}
\usepackage[dvips]{graphicx}
\usepackage{bm,latexsym,amsmath,amssymb,amsfonts}

\begin{document}

\title{A perturbative no-hair of form fields for higher dimensional static black holes}

\author{Tetsuya Shiromizu and Seiju Ohashi}
\affiliation{Department of Physics, Kyoto University, Kyoto 606-8502, Japan}

\author{Kentaro Tanabe}
\affiliation{Yukawa Institute for Theoretical Physics, Kyoto University, Kyoto 606-8502, Japan}

\begin{abstract}
In this paper we examine the static perturbation of $p$-form field strengths 
around higher dimensional Schwarzschild spacetimes. As a result, we can see that the 
static perturbations do not exist when $p\geq 3$. This result supports the 
no-hair properties of $p$-form fields. However, this does not exclude the 
presence of the black objects having non-spherical topology. 
\end{abstract}
\maketitle

\section{Introduction}

Motivated by the recent progress of superstring theory, higher 
dimensional black holes has been actively studied so far \cite{ER}. 
Different from the four dimensional cases, the conventional uniqueness 
theorem does not holds in stationary higher dimensional black holes. 
Indeed, there are several different black hole/ring spacetimes with same 
mass and angular momentum \cite{Myers:1986un, Emparan:2001wn}. See Ref. 
\cite{Emparan:2009cs} for a new approach ``blackfolds". 
But, if one considers static (electro)vacuum 
cases, the uniqueness theorem holds \cite{GIS1,GIS2} 
(See also Refs. \cite{Rogatko1,BM}) and then 
the spacetimes is the Schwarzschild-Tangherlini solution \cite{Sch} 
(The higher dimensional Reissner-Nordstr\"{o}m solution in electrovacuum 
cases). However, there are open questions even 
in static cases. If one puts other matter fields, it becomes difficult to show the uniqueness in general(See also 
Refs. \cite{Bekenstein, Shiromizu:2006vh}). 
For example, one might be 
interested in the higher form fields(say, $p$-form field strengths). 
According to the recent work \cite{EOS}, 
one can show the no-hair theorem for the cases with $ (n+1)/2 \leq p \leq (n-1) $ 
in $n$-dimensional asymptotically flat spacetimes. Note that the Maxwell 
field($p=2$) is out of the condition on $p$ and consistent with the 
presence of the Maxwell hair of charged black holes. However, there 
is a mystery about the presence of the hairs for $2 < p < (n+1)/2$. 
We should also note that the cases with $p\geq 3$ cannot have the conserved 
charge associated with $H_{(p)}$. Therefore, we intuitively guess that 
the monopole component of $H_{(p)}$ does not exist. In stationary cases, 
there is the exact solution with dipole hair \cite{Emparan:2004wy}.

In this paper, using the perturbation analysis, 
we will consider the possibility of the back hole spacetime with 
non-trivial $p$-form field strength hair. Since the background 
spacetimes are vacuum one, 
the $p$-form field perturbations are decoupled with the metric 
perturbations. So this set-up makes the analysis much easier 
than the cases of the perturbation analysis of ``charged" black holes. 
The analysis will 
show us that the  static perturbations of $p$-form field strength 
around the Schwarzschild-Tangherlini 
spacetime does not exist \footnote{After submission to ArXiv, we were informed that 
Ref. \cite{Gueven:1989wm} discussed the same issue. However, our study 
completes the analysis in the explicit way and for different cases with cosmological 
constant, and we also have a new implication by the consideration in Appendix.}. 
See the study on stationary metric 
perturbation for 
the Schwarzschild-Tangherlini spacetimes \cite{Kodama}. Our result suggests that the deformed black holes with spherical 
topology do not exist. But, this does not exclude the presence of the 
black hole solution with non-spherical topology. As the discussion in the 
appendix, if the solution exists, it seems to have both of the electric and 
magnetic hair of $p$-form field strengths simultaneously. 

The rest of this paper is organized as follows. In Sec. II, we 
describe the 
model, boundary conditions and hyperspherical harmonic functions(harmonics 
for the brevity). In Sec. III, we analyse the Maxwell fields from the 
pedagogical point of view. Then, in Sec. IV, we will discuss the static 
perturbation of general form field and show that there are no regular 
solutions. In Sec. V, we also have a little consideration of no-hair in 
asymptotically (anti)deSitter spacetimes. 
Finally we will summarise our work and discuss future issues in 
Sec. VI. In the appendix, we try to show the no-hair theorem in the cases 
with both of electric and magnetic parts of $p$-form field strength. 
But, we fail to do it.

\section{Field equations, boundary conditions and harmonics}

\subsection{Model}

We consider the system described by the Lagrangian 
%
\begin{equation}
{\cal L}=R-\frac{1}{p!}H_{(p)}^2,
\end{equation} 
%
where $R$ is the $n$-dimensional Ricci scalar and $H_{(p)}$ is the $p$-form 
field strength. $H_{(p)}$ has the $(p-1)$-form field potential as 
%
\begin{equation}
H_{(p)}=dB_{(p-1)}.
\end{equation} 
%
The field equations are 
%
\begin{equation}
R_{\mu\nu}=\frac{1}{p!}\Bigl(pH_\mu^{~\rho_1 \cdots \rho_{p-1}}H_{\nu \rho_1 \cdots \rho_{p-1}}
-\frac{p-1}{n-2}g_{\mu\nu}H_{(p)}^2 \Bigr)
\end{equation} 
%
and
%
\begin{equation}
\nabla_\mu H^\mu_{~\nu_1 \nu_2 \cdots \nu_{p-1}}=0,
\end{equation} 
%
where $\nabla_\mu$ is the covariant derivative with respect to 
$g_{\mu\nu}$. 
As in Ref. \cite{EOS}, we can include the dilation field too. 
However, the effect from the dilaton does not affect our result. 
Then, for simplicity, we will not include the dilaton fields in this study. 

\subsection{Boundary conditions}

Let us consider the boundary conditions. In general, the metric of static 
spacetimes can be written as  
%
\begin{equation}
ds^2=g_{\mu\nu}dx^\mu dx^\nu=-V^2(x^i)dt^2+g_{ij}(x^k)dx^i dx^j, \label{static}
\end{equation} 
%
where $x^i$ are spatial coordinate and $t$ is time coordinate. 
Since we mainly focus on asymptotically flat spacetimes, 
we suppose that the asymptotic boundary conditions are given by 
%
\begin{eqnarray}
& & V=1-\frac{m}{r^{n-3}}+O(1/r^{n-2}) \nonumber \\
& & g_{ij}=\Bigl(1+\frac{2}{n-3}\frac{m}{r^{n-3}} \Bigr) \delta_{ij}+O(1/r^{n-2}),
\end{eqnarray} 
%
where $m$ is the ADM mass. 
We will not use the aboves directly. 
From the asymptotic flatness, $H_{(p)}$ should decay at the infinity. 
Although we mainly discuss the asymptotically flat cases, we will 
address the no-hair of $H_{(p)}$ in asymptotically (anti)deSitter spacetimes 
shortly. 

The boundary condition on the event horizon $V=0$ comes from the regularity. 
To see this, we compute the Kretschmann invariant 
%
\begin{eqnarray}
R_{\mu\nu\rho\sigma}R^{\mu\nu\rho\sigma}
& = & 4R_{i0j0}R^{i0j0}+R_{ijkl}R^{ijkl} \nonumber \\
& = & \frac{4}{V^2}D_i D_j V D^i D^j V+R_{ijkl}R^{ijkl} \nonumber \\
& = & \frac{4}{V^2}
\Biggl[ \frac{1}{\rho^2}k_{ij}k^{ij}
+\frac{1}{\rho^4}(n^{i}\partial_{i} \rho)^2
+\frac{2}{\rho^4}({\cal D} \rho)^2\Biggr] \nonumber \\
& & +R_{ijkl}R^{ijkl}
\nonumber \\
& = & 
\frac{4}{V^2}
\Biggl[ \frac{1}{\rho^2}k_{ij}k^{ij}
+\frac{1}{\rho^2}(k-\rho D^2V)^2 \nonumber \\
& & +\frac{2}{\rho^4}({\cal D} \rho)^2\Biggr]
+R_{ijkl}R^{ijkl}, \label{ki}
\end{eqnarray} 
%
where we used $R_{i0j0}=VD_i D_j V$ in the second line and $D_i$ is the 
covariant derivative with respect to $g_{ij}$. In the third line 
$k_{ij}=h_{i}^{k} D_{k}n_{j}$ with $n_{i}=\rho D_{i}V$.
For the last line, we may be going to use the Einstein equation
%
\begin{eqnarray}
R_{00}& = & VD^2V \nonumber \\
& = & \frac{n-p-1}{(n-2)(p-1)!}H_0^{~i_1 \cdots i_{p-1}}H_{0 i_1 \cdots i_{p-1}}
\nonumber \\
& & +\frac{p-1}{(n-2)p!}V^2H_{i_1\cdots i_{p}}H^{i_1\cdots i_{p}}.
\label{r00}
\end{eqnarray} 
%
${\cal D}_i$ is the covariant derivative with respect to the induced metric 
$h_{ij}=g_{ij}-n_{i}n_{j}$. 

Thus, from Eqs. (\ref{ki}) and (\ref{r00}), the regularity implies 
%
\begin{equation}
k_{ij}|_{V=0}={\cal D}\rho|_{V=0}=0
\end{equation} 
%
%
\begin{eqnarray}
H_0^{~i_1 \cdots i_{p-1}}H_{0i_1 \cdots i_{p-1}}|_{V=0}=O(V^2). 
\end{eqnarray} 
%
and
%
\begin{eqnarray}
H^{i_1 \cdots i_{p}}H_{i_1 \cdots i_{p}}|_{V=0}=O(1). 
\end{eqnarray} 
%

In this paper we focus on the static perturbation around vacuum and spherical 
symmetric solutions. That is, the background $p$-form field does not exist. 
The background metric is given by 
%
\begin{equation}
ds^2_0=-f(r)dt^2+f(r)^{-1}dr^2+r^2d\Omega_{n-2}^2,
\end{equation} 
%
where $f(r)=1-(r_0/r)^{n-3}$ and $d\Omega_{n-2}^2=:\sigma_{AB}dx^Adx^B$ is 
the metric of the $(n-2)$-dimensional unit sphere.  
In this specific form, 
the static perturbation should satisfy 
%
\begin{eqnarray}
& & H_{0rA_1 \cdots A_{p-2}}|_{V=0}=O(1) \\
& & H_{0A_1 \cdots A_{p-1}}|_{V=0}=O(V)=O({\sqrt {f}}) \\
& & H_{rA_1\cdots A_{p-1}}|_{V=0}=O(V^{-1})=O(1/{\sqrt {f}})\\
& & H_{A_1 \cdots A_{p}}|_{V=0}=O(1).
\end{eqnarray} 
%

\subsection{Hyperspherical harmonic functions}

Since the background spacetimes has spherical symmetry, we 
can decompose all quantities in terms of hyperspherical 
harmonics defined on 
the sphere $S^{n-2}$ \cite{Gibbons, Rubin, Tanii}. In general, 
there are three type of harmonics, that is, 
scalar, vector and tensor types. The scalar harmonic function $Y$ follows 
%
\begin{eqnarray}
{\cal D}^2Y=-\ell(\ell+n-3) Y.
\end{eqnarray} 
%
The vector harmonic function $V_A$ satisfies 
%
\begin{eqnarray}
& & {\cal D}^A V_A=0 \\
& & {\cal D}^2V_A=-[\ell(\ell+n-3)-1]V_A.
\end{eqnarray} 
%
Since quantities which we will consider are often asymmetric tensor, 
we consider the totally anti-symmetric tensor harmonic function only 
%
\begin{eqnarray}
& & {\cal D}^{A_1}T_{A_1 \cdots A_{q}} =0 \\
& & {\cal D}^2T_{A_1 \cdots A_{q}}   =-[\ell(\ell+n-3)-q]
T_{A_1 \cdots A_{q}}. 
\end{eqnarray} 
%

Note that the static perturbation of metric and $p$-form fields are decoupled 
each others. This is due to the non-presence of the background field of 
$p$-form fields. Since we know that the possible static perturbation 
of the metric are 
$\ell=0,1$ modes. So if the mass is fixed, $\ell=0$ static modes vanishes. 
The $\ell=1$ modes can be absorbed to the redefinition of the coordinate, 
that is, it corresponds to the choice of the ``center" of the coordinate. 
Therefore, we will not consider the static perturbation of the metric.

\section{Maxwell fields}

As a pedagogical exercise, we will first consider the Maxwell fields. 
As already known, the uniqueness theorem of charged black hole 
(the higher dimensional Reissner-Nordstr\"{o}m solution) holds in this system. 
Therefore, the static {\it monopole} perturbation is only permitted. 
We will confirm this fact in this section. See Refs. \cite{old, KI} for the 
perturbation analysis of Reissner-Nordstr\"{o}m spacetimes. 

The each components of the Maxwell equation become
%
\begin{eqnarray}
\partial_r F_{rt}+\frac{n-2}{r}F_{rt}+\frac{1}{r^2f}{\cal D}^AF_{At}=0
\end{eqnarray} 
%
%
\begin{eqnarray}
{\cal D}^AF_{Ar}=0 \label{max-r}
\end{eqnarray} 
%
and
%
\begin{equation}
\partial_r F^r_{~A}+\frac{n-4}{r}F^r_{~A}+\frac{1}{r^2}{\cal D}^BF_{BA}=0.
\end{equation} 
%

\subsection{Gauge conditions}

We employ the following gauge
%
\begin{equation}
A_r={\cal D}^A A_A=0. \label{max-gauge}
\end{equation} 
%
This can been achieved by following standard argument. 
There is the gauge freedom of $A_\mu \to \tilde A_\mu=A_\mu +\partial_\mu \chi $
Then if we choose $\chi$ as 
%
\begin{eqnarray}
\chi=- \int dr A_r(r,x^A)+\eta(x^A) 
\end{eqnarray} 
%
we can set 
%
\begin{eqnarray}
A_r=0.
\end{eqnarray} 
%
Eq. (\ref{max-r}) implies 
%
\begin{eqnarray}
{\cal D}^2A_r-\partial_r ({\cal D}^AA_A)=0,
\end{eqnarray} 
%
and then 
%
\begin{eqnarray}
\partial_r ({\cal D}^AA_A)=0
\end{eqnarray} 
%
that is, ${\cal D}^AA_A$ does not depends on the coordinate of $r$. 
Using the remaining gauge freedom of $A_A \to \tilde A_A=A_A+\partial_A \eta (x^B)$ 
satisfying 
%
\begin{eqnarray}
{\cal D}^2\eta=-{\cal D}^A A_A,
\end{eqnarray} 
%
we can set 
%
\begin{eqnarray}
{\cal D}^A A_A=0.
\end{eqnarray} 
%

\subsection{Solutions}

Under the gauge condition of Eq. (\ref{max-gauge}), 
the Maxwell equation becomes 
%
\begin{eqnarray}
\partial_r^2 A_t+\frac{n-2}{r}\partial_r A_t +\frac{1}{r^2f}{\cal D}^2A_t=0
\end{eqnarray} 
%
and
%
\begin{equation}
\partial_r^2A_A+\Bigl(\frac{n-4}{r}+\frac{f'}{f}  \Bigr)\partial_r A_A+\frac{{\cal D}^2-(n-3)}{r^2f}A_A=0.
\end{equation} 
%
Here we expand $A_t, A_A$ in terms of harmonics as
%
\begin{equation}
A_t=G(r)Y,~~A_A=H(r)V_A.
\end{equation} 
%

Let us first solve the equation for $A_t$. Introducing the new variable 
$x$ defined by 
%
\begin{equation}
x:=\Bigl( \frac{r_0}{r} \Bigr)^{n-3},
\end{equation} 
%
the solution can be written in the analytic form of 
%
\begin{equation}
G(r)=Br^{-(n+\ell-3)}F(\alpha,\beta,\gamma;x)+Cr^\ell F(\alpha',\beta',\gamma';x),
\end{equation} 
%
where $F(\alpha,\beta,\gamma;x)$ is the hypergeometric function, and 
%
\begin{eqnarray}
& & \alpha=\frac{\ell}{n-3}\\
& & \beta=\frac{n+\ell-3}{n-3}{}\\
& & \gamma=\frac{2(n+\ell-3)}{n-3}=\alpha+\beta+1
\end{eqnarray} 
%
and
%
\begin{eqnarray}
& & \alpha'=-\frac{\ell}{n-3}\\
& & \beta'=-\frac{n+\ell-3}{n-3}\\
& & \gamma'=-\frac{2\ell}{n-3}=\alpha'+\beta'+1.
\end{eqnarray} 
%
From the asymptotic flatness, we must set $C=0$ and 
the solution becomes  
%
\begin{equation}
A_t=Br^{-(n+\ell-3)}F(\alpha,\beta,\gamma;x)Y. 
\end{equation} 
%
Now we compute the field strength 
%
\begin{eqnarray}
F_{r0} & = & -(n+\ell-3)r^{-(n+\ell-2)}F(\alpha,\beta,\gamma;x)Y \nonumber \\
& & -(n-3)r^{-(n+\ell-3)}\frac{x}{r} \frac{d}{dx}F(\alpha,\beta,\gamma;x)Y \nonumber \\
& = & -(n+\ell-3)r^{-(n+\ell-2)}F(\alpha,\beta,\gamma;x)Y \nonumber \\
& & -(n-3)\frac{\alpha \beta}{\gamma}r^{-(n+\ell-3)}\frac{x}{r} \times \nonumber \\
& & ~~~~ \frac{d}{dx}F(\alpha+1,\beta+1,\gamma+1;x)Y
\end{eqnarray} 
%
Let us examine the behavior on the horizon. Since 
%
\begin{equation}
F(\alpha+1,\beta+1,\gamma+1;1)
=\frac{\Gamma(\beta+1)\Gamma(\gamma-\alpha-\beta-1)}{\Gamma(\gamma-\alpha)\Gamma(\gamma-\beta)}
\label{hgf}
\end{equation} 
%
and $\Gamma(\gamma-\alpha-\beta-1)=\Gamma(0)$, it diverges except for $\ell \neq 0$. 
This means that the second term in the right-hand side of Eq. (\ref{hgf}) 
diverges. 
The case of $\ell=0$ is special. In this case, $\alpha$ vanishes and then the 
second term disappears and the solution will be regular everywhere outside of 
the black holes. Thus, 
the monopole component($\ell=0$) is only permitted. Of course, this is the 
case of the Reissner-Nordstr\"{o}m solution. 

Next we solve the equation for $A_A$ and then we have 
the analytic solution as  
%
\begin{equation}
H(r)=Br^{-(\ell+n-4)}F(\alpha,\beta,\gamma;x)+Cr^{\ell+1}F(\alpha',\beta',\gamma';x),
\end{equation}
%
where 
%
\begin{eqnarray}
& & \alpha = \frac{\ell+n-4}{n-3}\\
& & \beta=\frac{\ell+n-2}{n-3}\\
& & \gamma=2\frac{\ell+n-3}{n-3}=\alpha+\beta
\end{eqnarray}
%
and
%
\begin{eqnarray}
& & \alpha'=-\frac{\ell+1}{n-3}\\
& & \beta'=-\frac{\ell-1}{n-3}\\
& & \gamma'=-\frac{2\ell}{n-3}.
\end{eqnarray} 
%
From the asymptotic flatness, we must set $C=0$ and 
the solution becomes 
%
\begin{equation}
A_A=Br^{-(\ell+n-4)}F(\alpha,\beta,\gamma;x)V_A.
\end{equation}
%
Since $\gamma=\alpha+\beta$, on the event horizon, 
%
\begin{equation}
F(\alpha,\beta,\gamma;1)=\frac{\Gamma(\gamma)\Gamma(0)}{\Gamma(\gamma-\alpha)\Gamma(\gamma-\beta)}
\end{equation}
%
diverges. Therefore there is no regular solution. 

As a conclusion, the regular solution is $\ell=0$ mode only of $A_t$ 
which corresponds to the Reissner-Nordstr\"{o}m solution. This is 
well-known fact. 

\section{Higher form fields}

In this section we examine the static perturbation of $H_{(p)}$ fields 
with $p \geq 3$. The field equations are 
%
\begin{equation}
{\cal D}^A H_{AtrA_1A_2\cdots A_{p-3}}=0 \label{HtrA}
\end{equation}
%
%
\begin{eqnarray}
& & \partial_r H_{rtA_1A_2 \cdots A_{p-2}}+\frac{n-2(p-1)}{r}H_{rtA_1A_2 \cdots A_{p-2}}
\nonumber \\
& & +\frac{1}{r^2f}{\cal D}^BH_{BtA_1\cdots A_{p-2}}=0\label{HtA}
\end{eqnarray} 
%
%
\begin{equation}
{\cal D}^A H_{ArA_1\cdots A_{p-2}}=0.\label{HrA}
\end{equation} 
%
and
%
\begin{eqnarray}
& & \partial_r H_{rA_1 \cdots A_{p-1}}+\Bigl(\frac{n-2p}{r}+\frac{f'}{f} \Bigr)H_{rA_1 \cdots A_{p-1}}
\nonumber \\
& & +\frac{1}{r^2f}{\cal D}^BH_{BA_1\cdots A_{p-1}}=0.\label{HAB}
\end{eqnarray} 
%

\subsection{Gauge conditions}

Using the gauge freedom of $B_{\mu_1 \cdots \mu_{p-1}}\to \tilde B_{\mu_1 \cdots \mu_{p-1}}=
B_{\mu_1 \cdots \mu_{p-1}}+\partial_{[\mu_1} C_{\mu_2 \cdots \mu_{p-1}]}$, we 
can show that one can choose the following gauge condition
%
\begin{eqnarray}
& & {\cal D}^A B_{tAA_1\cdots A_{p-3}}=0 \label{bg1} \label{g-bta} \\ 
& & {\cal D}^A B_{rAA_1\cdots A_{p-3}}=0\label{bg2}\\
& & {\cal D}^B B_{BA_1\cdots A_{p-2}}=0\label{bg3}.
\end{eqnarray} 
%
With Eq. (\ref{g-bta}), the field equation shows 
%
\begin{eqnarray}
B_{trA_1\cdots A_{p-3}}=0.\label{gb4}
\end{eqnarray} 
%
The detail can been seen by the following argument. 
The gauge transformation gives us 
%
\begin{eqnarray}
& & {\cal D}^A \tilde B_{tAA_1 \cdots A_{p-3}} \nonumber \\
& &~~=  {\cal D}^A B_{tAA_1 \cdots A_{p-3}}\nonumber \\
& & ~~~~-[{\cal D}^2-(p-3)(n-p+1)]C_{tA_1\cdots A_{p-3}}, 
\end{eqnarray} 
%
where we already imposed  ${\cal D}^{A_1} C_{tA_1\cdots A_{p-3}}=0$. 
Then we take $C_{tA_1\cdots A_{p-3}}$ satisfying 
%
\begin{eqnarray}
[{\cal D}^2-(p-3)(n-p+1)]C_{tA_1\cdots A_{p-3}}={\cal D}^A \tilde B_{tAA_1 \cdots A_{p-3}}
\nonumber \\
\end{eqnarray} 
%
Note that there exists the solutions for $C_{tA_1\cdots A_{p-3}}$. 
Then this implies 
%
\begin{eqnarray}
& & {\cal D}^A B_{tAA_1\cdots A_{p-3}}=0
\end{eqnarray} 
%
In this case, Eq. (\ref{HtrA}) becomes 
%
\begin{eqnarray}
[{\cal D}^2-(p-3)(n-p-1)]B_{trA_1 \cdots A_{p-3}}=0.
\end{eqnarray} 
%
In terms of harmonics, $B_{trA_1 \cdots A_{p-3}}$ will be expanded as
%
\begin{eqnarray}
B_{trA_1 \cdots A_{p-3}}=J(r)T_{A_1 \cdots A_{p-3}}. 
\end{eqnarray} 
%
Then 
%
\begin{eqnarray}
(\ell+p-3)(\ell+n-p) J(r)=0.
\end{eqnarray} 
%
Except for the special case with $\ell=0, p=3$, it is easy to see that 
%
\begin{eqnarray}
J(r)=0
\end{eqnarray} 
%
holds. We can also show $J(r)=0$ even for $\ell=0, p=3$ case by a 
distinct argument. In fact, we can use the remaining gauge freedom of 
$\tilde B_{tr}=B_{tr}-\partial_r C_t(r)$. Then, taking 
%
\begin{eqnarray}
C_t(r)=\int^r dr B_{tr}(r),
\end{eqnarray} 
%
we can set 
%
\begin{eqnarray}
B_{tr}(r)=0. 
\end{eqnarray} 
%
Therefore, without loss of generality, we can conclude that 
%
\begin{eqnarray}
B_{trA_1 \cdots A_{p-3}}=0
\end{eqnarray} 
%
holds. 

Next we will ask if we can take the  gauge condition of Eq. (\ref{bg2}). 
To see this, we first look at 
%
\begin{eqnarray}
& &{\cal D}^A \tilde B_{rAA_1\cdots A_{p-3}} \nonumber \\
& & ~~=  {\cal D}^A B_{rAA_1\cdots A_{p-3}}+\partial_r({\cal D}^AC_{AA_1\cdots A_{p-3}})
\nonumber \\
& & ~~~~-[{\cal D}^2-(p-3)(n-p+1)]C_{rA_1\cdots A_{p-3}},
\end{eqnarray} 
%
where we imposed ${\cal D}^AC_{rAA_1 \cdots A_{p-3}}=0$. 

Using $C_{\mu_1 \cdots \mu_{p-2}}$ satisfying 
%
\begin{eqnarray}
& &\partial_r({\cal D}^AC_{AA_1\cdots A_{p-3}})=0 \\
& & [{\cal D}^2-(p-3)(n-p+1)]C_{rA_1\cdots A_{p-3}} \nonumber \\
& & ~~~~~~~~~~={\cal D}^A B_{rAA_1\cdots A_{p-3}},
\end{eqnarray} 
%
we can set 
%
\begin{eqnarray}
{\cal D}^A B_{rAA_1\cdots A_{p-3}}=0.
\end{eqnarray} 
%

Finally we consider the following transformation
%
\begin{eqnarray}
& & {\cal D}^A \tilde B_{AA_1\cdots A_{p-2}} \nonumber \\
& & ~~ =  {\cal D}^A B_{AA_1\cdots A_{p-2}}\nonumber \\
& & ~~~~+
[{\cal D}^2-(n-p)(p-2)]C_{A_1\cdots A_{p-2}},
\end{eqnarray} 
%
where we imposed ${\cal D}^AC_{AA_1\cdots A_{p-3}}=0$. 
Then, taking $C_{A_1\cdots A_{p-2}}$ satisfying 
%
\begin{eqnarray}
[{\cal D}^2-(n-p)(p-2)]C_{A_1\cdots A_{p-2}}=-{\cal D}^A B_{AA_1\cdots A_{p-2}},
\end{eqnarray} 
%
we can adopt the gauge of 
%
\begin{eqnarray}
{\cal D}^A B_{AA_1\cdots A_{p-2}}=0.
\end{eqnarray} 
%

\subsection{Solutions}

Now, under the current gauge conditions, Eq. (\ref{HtA}) becomes 
%
\begin{eqnarray}
& & \partial_r^2 B_{tA_1\cdots A_{p-2}}+\frac{n-2(p-1)}{r}\partial_r B_{tA_1\cdots A_{p-2}}
\nonumber \\
& & +\frac{{\cal D}^2-(n-p)(p-2)}{r^2f}B_{tA_1\cdots A_{p-2}}=0.
\end{eqnarray} 
%
Here we expand $B_{tA_1\cdots A_{p-2}}$ in terms of the harmonics as
%
\begin{eqnarray}
B_{tA_1\cdots A_{p-2}}=K(r)T_{A_1\cdots A_{p-2}}.
\end{eqnarray} 
%
Then the above equation becomes 
%
\begin{eqnarray}
& & \partial_r^2K+\frac{n-2(p-1)}{r}\partial_rK \nonumber \\
& & ~~~~-\frac{(\ell+p-2)(\ell+n-p-1)}{r^2f}K=0. \label{Keq}
\end{eqnarray} 
%
The solution has been found in the analytic form of 
%
\begin{eqnarray}
K(r)& = & Br^{-(\ell+n-p-1)}F(\alpha,\beta,\gamma;x)\nonumber \\
& & +Cr^{\ell+p-2}
F(\alpha',\beta',\gamma';x),
\end{eqnarray} 
%
where 
%
\begin{eqnarray}
& & \alpha=\frac{\ell+p-2}{n-3}\\
& & \beta=\frac{\ell+n-p-1}{n-3}\\
& & \gamma=2\frac{\ell+n-3}{n-3}=\alpha+\beta+1
\end{eqnarray} 
%
and
%
\begin{eqnarray}
& & \alpha'=-\frac{\ell+p-2}{n-3}\\
& & \beta'=-\frac{\ell+n-p-1}{n-3}\\
& & \gamma'=-2\frac{\ell}{n-3}.
\end{eqnarray} 
%
From the asymptotic flatness, the solution will be 
%
\begin{eqnarray}
B_{tA_1\cdots A_{p-2}}=r^{-(\ell+n-p-1)}F(\alpha,\beta,\gamma;x)
T_{A_1\cdots A_{p-2}}. 
\end{eqnarray} 
%
Now we can compute the field strength $H_{rtA_1\cdots A_{p-2}}$
%
\begin{eqnarray}
& & H_{rtA_1\cdots A_{p-2}} \nonumber \\
& &~~=  -(\ell+n-p-1)r^{-(\ell+n-p)}F(\alpha,\beta,\gamma;x)
T_{A_1\cdots A_{p-2}}
\nonumber \\
& & ~~~~~-(n-3)\frac{\alpha\beta}{\gamma}\frac{x}{r}r^{-(\ell+n-p-1)} \times \nonumber \\
& & ~~~~~~~~F(\alpha+1,\beta+1,\gamma+1;x)T_{A_1\cdots A_{p-2}}. \label{hgf2} \\
\end{eqnarray} 
%
Since 
%
\begin{eqnarray}
F(\alpha+1,\beta+1,\gamma+1;1)
=\frac{\Gamma(\gamma+1)\Gamma(\gamma-\alpha-\beta-1)}{\Gamma(\gamma-\alpha)\Gamma(\gamma-\beta)},
\nonumber \\
\end{eqnarray} 
%
and $\Gamma(\gamma-\alpha-\beta-1)=\Gamma(0)=-\infty$, the second term 
in the right-hand side of Eq. (\ref{hgf2}) diverges at the horizon. 
Thus, there are no regular solutions. 

In the current gauge, Eq. (\ref{HAB}) becomes 
%
\begin{eqnarray}
& & \partial_r^2 B_{A_1\cdots A_{p-1}}+\Bigl(\frac{n-2p}{r}+\frac{f'}{f}\Bigr)
\partial_r B_{A_1\cdots A_{p-1}} \nonumber \\
& & ~~+\frac{{\cal D}^2-(n-p-1)(p-1)}{r^2f}B_{A_1\cdots A_{p-1}}=0.
\label{eq-bab}
\end{eqnarray} 
%
Let us expand $B_{A_1 \cdots A_{p-1}}$ in terms of harmonics as 
%
\begin{eqnarray}
B_{A_1 \cdots A_{p-1}}=L(r)T_{A_1 \cdots A_{p-1}}.
\end{eqnarray} 
%
Then Eq. (\ref{eq-bab}) becomes 
%
\begin{eqnarray}
& & \partial_r^2 L+\Bigl(\frac{n-2p}{r}+\frac{f'}{f}\Bigr)
\partial_r L \nonumber \\
& & -\frac{(\ell+p-1)(\ell+n-p-2)}{r^2f}L=0. \label{Leq}
\end{eqnarray} 
%

The solution is given by 
%
\begin{eqnarray}
L(r) & = & Br^{-(\ell+n-p-2)}F(\alpha,\beta,\gamma;x)
\nonumber \\
& & +Cr^{\ell+p-1}F(\alpha',\beta',\gamma';x),
\end{eqnarray} 
%
where 
%
\begin{eqnarray}
& & \alpha=\frac{\ell+n-p-2}{n-3}\\
& & \beta=\frac{\ell+n+p-4}{n-3}\\
& & \gamma=2\frac{n+\ell-3}{n-3}=\alpha+\beta
\end{eqnarray} 
%
and
%
\begin{eqnarray}
& & \alpha'=-\frac{\ell+p-1}{n-3}\\
& & \beta'=-\frac{\ell-p+1}{n-3}\\
& & \gamma'=-2\frac{\ell}{n-3}=\alpha'+\beta'.
\end{eqnarray} 
%
The asymptotic flatness implies $C=0$ and then we see 
%
\begin{eqnarray}
B_{A_1\cdots A_{p-1}}=Br^{-(\ell+n-p-2)}F(\alpha,\beta,\gamma;x)
T_{A_1 \cdots A_{p-1}}.
\end{eqnarray} 
%
Since
%
\begin{eqnarray}
F(\alpha,\beta,\gamma;1)=\frac{\Gamma(\gamma)\Gamma(0)}{\Gamma(\gamma-\alpha)\Gamma(\gamma-\beta)},
\end{eqnarray} 
%
we see the singular behaviors of the field strength as 
%
\begin{eqnarray}
H_{rA_1\cdots A_{p-1}}=O(1/(r-r_0))
\end{eqnarray} 
%
and
%
\begin{eqnarray}
H_{rA_1\cdots A_{p-1}}=O(1/(r-r_0)).
\end{eqnarray} 
%
As a conclusion we can show that black holes cannot have the hair of the 
$p$-form field strengths.

\section{Asymptotically (anti)deSitter spacetimes}

So far we concentrated on asymptotically flat spacetimes and 
could have the analytic solution for the equation of static perturbation. 
On the other hand, this is not the case once one 
turns on the cosmological constant. Without solving the equations, however, 
we can ask if the solution exists. Note that the equations for the static 
perturbations does not changed except for the expression of $f(r)$ in the 
metric of the background spacetimes. 

\subsection{deSitter cases}

First we consider the cases with positive cosmological constant. 
In this case, the background spacetime is higher dimensional 
Schwarzschild-deSitter spacetime and $f(r)$ in the metric becomes 
%
\begin{eqnarray}
f(r)=1-(r_0/r)^{n-3}-r^2/a^2.
\end{eqnarray} 
%
Under a certain cases with parameter $r_0$ and $a$, there are two horizons, 
black hole and cosmological horizons at $r_h$ and $r_c$. Note 
that $r_c>r_h$. 

From Eq. (\ref{Keq}), we have the following relation
%
\begin{eqnarray}
& & \int_{r_h}^{r_c} dr r^{n-2(p-1)}\Biggl[(\partial_r K)^2
\nonumber \\
& & ~~+\frac{(\ell+p-2)(\ell+n-p-1)}{r^2f}K^2  \Biggl]
\nonumber \\
& & ~~~~=\Bigl[r^{n-2(p-1)}K\partial_rK \Bigr]^{r_c}_{r_h}
\end{eqnarray} 
%
Since the presence of the cosmological constant does not disturb the behavior the 
horizons, the same regularity conditions are imposed on the both of the horizons. 
Thus we can see that the boundary term in the right-hand side vanishes and 
then 
%
\begin{eqnarray}
K=0.
\end{eqnarray} 
%

In the same way, from Eq. (\ref{Leq}), we have 
%
\begin{eqnarray}
& & \int_{r_h}^{r_c} dr r^{n-2p}\Biggl[f(\partial_r L)^2 \nonumber \\
& & ~~+\frac{(\ell+p-1)(\ell+n-p-2)}{r^2}L^2  \Biggl]
\nonumber \\
& & ~~~~=\Bigl[r^{n-2p}fL\partial_rL \Bigr]^{r_c}_{r_h}.
\end{eqnarray} 
%
Since the boundary term vanishes, we can see that 
%
\begin{eqnarray}
L=0
\end{eqnarray} 
%
holds. Therefore, there are no regular static perturbation in the region 
of $r_h \leq r \leq r_c$. 

\subsection{anti-deSitter cases}

Next let us consider asymptotically anti-deSitter cases. In this case, 
$f(r)=1-(r_0/r)^{n-3}+r^2/a^2$. Then, near the infinity, $K$ follows the equation approximately
%
\begin{eqnarray}
\partial_r^2 K+\frac{n-2(p-1)}{r}\partial_r K \simeq 0.
\end{eqnarray} 
%
Then the solution is approximately given by  
%
\begin{eqnarray}
K \simeq \frac{A}{r^{n-2p+1}}.
\end{eqnarray} 
%
In the current case, Eq. (\ref{Keq}) gives us 
%
\begin{eqnarray}
& & \int_{r_h}^{\infty} dr r^{n-2(p-1)}\Biggl[(\partial_r K)^2 \nonumber \\
& & ~~+\frac{(\ell+p-2)(\ell+n-p-1)}{r^2f}K^2  \Biggl]
\nonumber \\
& & ~~~~=\Bigl[r^{n-2(p-1)}K\partial_rK \Bigr]^{\infty}_{r_h}.
\end{eqnarray} 
%
Since the boundary term near the infinity is roughly estimated as 
$\int dr 1/r^{n-2p+2} $, one has to impose $(n+1)/2 > p$ in order to make 
it finite. Thus, if we impose $(n+1)/2 > p$, the boundary term vanishes and then 
we can conclude 
%
\begin{eqnarray}
K=0.
\end{eqnarray} 
%
Similar result will be obtain for $L$, that is, $L=0$. 

As a consequence, we can see that there no static perturbation of 
$p$-form field strength in asymptotically anti-deSitter spacetimes too.  

\section{Summary and discussions}

In this paper we studied the static perturbation of $p$-form field 
strengths 
for the Schwarzschild-Tangherlini spacetime and then we could show that 
the black holes cannot have $p$-form hair except for the Maxwell 
cases($p=2, \ell=0$). This work is initiated by remaining issue in 
the no-hair theorem \cite{EOS} of $p$-form fields in higher dimensional 
black hole spacetimes. That is, therein, there is a limitation of $p$ as 
$p\geq (n+1)/2$ in the proof of the no-hair theorem. Therefore, it was natural to ask if 
the no-hair properties with $p<(n+1)/2$ is. Our current result supports 
no-hair properties of $p$-form field strength with $p\geq 3$ regardless 
of such limitation. 

Our analysis is based on the perturbation and then the topology of 
black holes is limited to be sphere. So if one thinks of another 
topology like black ring, there are still possibility to have a 
solution. According to the appendix, however, the solution may have the 
both electric and magnetic hairs if it exists. They will be addressed in near 
future study.

\begin{acknowledgments}
 TS is partially supported by
Grant-Aid for Scientific Research from Ministry of Education, Science,
Sports and Culture of Japan (Nos.~21244033,~21111006,~20540258 and 19GS0219).
SO thanks Professor Takashi Nakamura for his continuous encouragement.
KT is supported by JSPS Grant-Aid for Scientific Research (No.~21-2105).
SO and KT are supported by the Grant-in-Aid for the Global 
COE Program ``The Next Generation of Physics, Spun from Universality 
and Emergence'' from the Ministry of Education, Culture, Sports, Science 
and Technology (MEXT) of Japan.  
\end{acknowledgments}

\appendix

\section{no-dipole-hair theorem revisit} 

In this appendix, we revisit the no-hair theorem of $p$-form fields 
strengths 
in static asymptotically flat spacetimes \cite{EOS}. 
In the theorem, one supposes the presence of the electric part only. 
Then, if $p\geq (n+1)/2 $, we can show that the $p$-form hair does not 
exist. From this, if one supposes the presence of the magnetic part of 
$p$-form field strengths only, we expect that similar theorem holds. 
In fact, its dual version is the electric part of $(n-p)$-form field 
strengths. Therefore, we would guess that, if $(n+1)/2 \geq (n-p)$, 
the magnetic parts of $p$-form field strengths does 
not exist. The above condition is rearranged as $p \geq (n-1)/2$. 
The above consideration indicates the breakdown of the proof of the 
no-hair of $p$-form field strengths if both parts exist. 
On the other hand, the argument in the main text 
indicates the no-hair of $p$-forms except for $p=2$. Or it may suggests 
the presence of the solutions which cannot be explained by the perturbation 
on the Schwarzschild spacetime. 

Let us examine if the no-hair theorem holds in the details. 
Here we includes the dilation to the system described by the Lagrangian
%
\begin{eqnarray}
{\cal L}=R-\frac{1}{2}(\nabla \phi)^2-\frac{1}{p!}H_{(p)}^2,
\end{eqnarray} 
%
where $\phi$ is the dilation field. The Einstein equation is 
%
\begin{eqnarray}
R_{\mu\nu}& = & \frac{1}{2}\nabla_\mu \phi \nabla_\nu \phi 
+\frac{1}{p!}e^{-\alpha\phi}
\Biggl(pH_\mu^{~\alpha_1 \cdots \alpha_{p-1}}H_{\nu\alpha_1 \cdots \alpha_{p-1}} 
\nonumber \\
& & -\frac{p-1}{n-2}g_{\mu\nu}H_{(p)}^2 \Biggr).
\end{eqnarray} 
%
Since we will not the equations for the $p$-form fields and dilation, we do not write down these equations. 
Different from Ref. \cite{EOS}, 
we will not assume that the $p$-form fields have the electric components 
only. The metric of static spacetimes is written as Eq. (\ref{static}). 
From the Einstein equation, then, we can see 
%
\begin{eqnarray}
R_{00}& = & VD^2V \nonumber \\
& = & \frac{n-p-1}{(n-2)(p-1)!}
e^{-\alpha \phi}
H_0^{~i_1\cdots i_{p-1}}H_{0i_1\cdots i_{p-1}}
\nonumber \\
& & +\frac{p-1}{(n-2)p!}V^2e^{-\alpha \phi}
H_{i_1\cdots i_p}H^{i_1\cdots i_p}
\end{eqnarray} 
%
and
%
\begin{eqnarray}
R_{ij} & = & {}^{(n-1)}R_{ij}-\frac{1}{V}D_i D_j V
\nonumber \\
& = & \frac{1}{2}D_i \phi D_j \phi \nonumber \\
& & +\frac{1}{(p-2)!}e^{-\alpha\phi}
\Bigl(
H_i^{~0k_1 \cdots k_{p-2}}H_{j0k_1\cdots k_{p-2}}\nonumber \\
& & -\frac{1}{n-2}g_{ij}H_{0k_1\cdots k_{p-1}}H^{0k_1\cdots k_{p-1}}
\Bigr)\nonumber \\
& & +\frac{1}{(p-1)!}e^{-\alpha\phi}
\Bigl( 
H_i^{~k_1\cdots k_{p-1}}H_{jk_1\cdots k_{p-1}}
\nonumber \\
& & -\frac{p-1}{p(n-2)}g_{jk}H_{k_1\cdots k_{p-1}}
\Bigr)
\end{eqnarray} 
%
hold. Moreover, we can compute the Ricci scalar of $g_{ij}$ as 
%
\begin{eqnarray}
{}^{(n-1)}R
& = & \frac{1}{2}(D\phi)^2
+\frac{1}{(p-1)!}\frac{e^{-\alpha\phi}}{V^2}H_0^{~i_1\cdots i_{p-1}}H_{0i_1\cdots i_{p-1}}
\nonumber \\
& & +\frac{1}{p!}e^{-\alpha\phi}
H_{i_1\cdots i_p}H^{i_1\cdots i_p}.
\end{eqnarray} 
%

The outline of the proof will be as follows if it works. We 
first consider the conformal transformation of $t=$constant 
hypersurfaces so that the Ricci scalar is non-negative and 
the ADM mass vanishes. Then we will apply the positive mass theorem 
\cite{PET,PET2} and then show that the conformally transformed spacetime is flat and 
the $p$-form hair does not exist. We know that the vacuum black hole 
spacetimes with 
conformally flat static slices must be spherical symmetry. Thus, the 
resulted spacetimes is the Schwarzschild spacetime. 

Let us look at the details. For the proof of no-hair, we will consider 
the two conformal transformations given by 
%
\begin{eqnarray}
\tilde g_{ij}^\pm=\Omega_\pm^2g_{ij},
\end{eqnarray} 
%
where 
%
\begin{eqnarray}
\Omega_\pm=\Bigl( \frac{1\pm V}{2}\Bigr)^{\frac{2}{n-3}}=:\omega^{\frac{2}{n-3}}_\pm.
\end{eqnarray} 
%
Then 
%
\begin{eqnarray}
& & \Omega^2_\pm{}^{(n-1)}\tilde R \nonumber \\
& & ~~={}^{(n-1)}R\mp \frac{2(n-2)}{n-3}\omega_\pm^{-1}D^2V 
\nonumber \\
& & ~~=\frac{1}{2}(D\phi)^2
+\frac{1}{(p-1)!}\frac{e^{-\alpha\phi}}{V^2}
\frac{\lambda_\pm}{\omega_\pm}H_0^{~i_1\cdots i_{p-1}}
H^{0i_1\cdots i_{p-1}}\nonumber \\
& & ~~~~+\frac{1}{p!}e^{-\alpha\phi}
\frac{\mu_\pm}{\omega_\pm}
H_{i_1\cdots i_p}H^{i_1\cdots i_p},
\end{eqnarray} 
%
where
%
\begin{eqnarray}
\lambda_\pm
=\frac{1\mp \frac{3n-4p-1}{n-3}}{2}
\end{eqnarray} 
%
and
%
\begin{eqnarray}
\mu_\pm
=\frac{1\pm \frac{n-4p+1}{n-3}}{2}.
\end{eqnarray} 
%
If ${}^{(n-1)}\tilde R_\pm \geq 0$, we can proceed the proof. 
However, we cannot do that.  The sufficient condition 
for ${}^{(n-1)}\tilde R_\pm \geq 0$ 
are $\lambda_\pm \geq 0$ and $\mu_\pm \geq 0$. Each conditions become
%
\begin{eqnarray}
p \geq \frac{n+1}{2}~~{\rm and}~~p\leq \frac{n-1}{2},
\end{eqnarray} 
%
respectively. The both conditions together do not hold manifestly. Therefore, 
we can say nothing about the no-hair for the cases having both of electric and magnetic 
$p$-form fields. This results may suggest the presence of the $p$-form hairy static 
black object solutions.




\begin{thebibliography}{99}


\bibitem{ER}
  R.~Emparan and H.~S.~Reall,
  Living Rev.\ Rel.\  {\bf 11}, 6 (2008)
  [arXiv:0801.3471 [hep-th]].

\bibitem{Myers:1986un}
  R.~C.~Myers and M.~J.~Perry,
  Annals Phys.\  {\bf 172}, 304 (1986).


\bibitem{Emparan:2001wn}
  R.~Emparan and H.~S.~Reall,
  Phys.\ Rev.\ Lett.\  {\bf 88}, 101101 (2002)
  [arXiv:hep-th/0110260].

\bibitem{Emparan:2009cs}
  R.~Emparan, T.~Harmark, V.~Niarchos and N.~A.~Obers,
  Phys.\ Rev.\ Lett.\  {\bf 102} 191301 (2009) 
  [arXiv:0902.0427 [hep-th]];
  J. High Energy Phys. 03 (2010) 063
  [arXiv:0910.1601 [hep-th]];
  M.~M.~Caldarelli, R.~Emparan and B.~Van Pol,
  arXiv:1012.4517 [hep-th].


\bibitem{GIS1}
  G.~W.~Gibbons, D.~Ida and T.~Shiromizu,
  Prog.\ Theor.\ Phys.\ Suppl.\  {\bf 148}, 284 (2002)
  [arXiv:gr-qc/0203004].


\bibitem{GIS2}
  G.~W.~Gibbons, D.~Ida and T.~Shiromizu,
  Phys.\ Rev.\ Lett.\  {\bf 89}, 041101 (2002)
  [arXiv:hep-th/0206049];
  Phys.\ Rev.\  D {\bf 66}, 044010 (2002)
  [arXiv:hep-th/0206136].


\bibitem{Rogatko1}
  M.~Rogatko,
  Phys.\ Rev.\  D {\bf 67}, 084025 (2003)
  [arXiv:hep-th/0302091],

\bibitem{BM}
G.L. Bunting and A.K.M. Masood-ul-Alam, Gen. Rel. Grav. {\bf 19}, 
147 (1987). 


 \bibitem{Sch}
  F.~R.~Tangherlini,
  Nuovo Cim.\  {\bf 27}, 636 (1963).

\bibitem{Bekenstein}
  J.~D.~Bekenstein,
  Phys.\ Rev.\  D {\bf 5}, 1239 (1972);
 Phys.\ Rev.\  D {\bf 5}, 2403 (1972).


\bibitem{Shiromizu:2006vh}
  T.~Shiromizu, S.~Yamada and H.~Yoshino,
  J.\ Math.\ Phys.\  {\bf 47}, 112502 (2006)
  [arXiv:gr-qc/0605029].

\bibitem{EOS}
  R.~Emparan, S.~Ohashi and T.~Shiromizu,
  Phys.\ Rev.\  D {\bf 82}, 084032 (2010)
  [arXiv:1007.3847 [hep-th]].

\bibitem{Emparan:2004wy}
  R.~Emparan,
  J. High Energy Phys. 03 (2004) 064
  [arXiv:hep-th/0402149].

\bibitem{Gueven:1989wm}
  R.~Gueven,
  Class.\ Quant.\ Grav.\  {\bf 6}, 1961 (1989).


\bibitem{Kodama}
  H.~Kodama,
  Prog.\ Theor.\ Phys.\  {\bf 112}, 249 (2004)
  [arXiv:hep-th/0403239].





\bibitem{Gibbons}
  G.~W.~Gibbons and M.~J.~Perry,
  Nucl.\ Phys.\  B {\bf 146}, 90 (1978).


\bibitem{Rubin}
  M.~A.~Rubin and C.~R.~Ordonez,
  J.\ Math.\ Phys.\  {\bf 25}, 2888(1984); {\bf 26}, 65 (1985).

\bibitem{Tanii}
  H.~Hata and Y.~Tanii,
  Nucl.\ Phys.\  B {\bf 624}, 283 (2002)
  [arXiv:hep-th/0110222].


\bibitem{old}
  V.~Moncrief,
  Phys.\ Rev.\  D {\bf 9}, 2707 (1974);
  F.~J.~Zerilli,
  Phys.\ Rev.\  D {\bf 9}, 860 (1974).

\bibitem{KI}
  H.~Kodama and A.~Ishibashi,
  Prog.\ Theor.\ Phys.\  {\bf 111}, 29 (2004)
  [arXiv:hep-th/0308128].
  
\bibitem{PET}
  E.~Witten,
  Commun.\ Math.\ Phys.\  {\bf 80}, 381 (1981),



\bibitem{PET2}
 R.~Schoen and S.~T.~Yau,
 Commun.\ Math.\ Phys.\  {\bf 65}, 45 (1979);
 R.~Schoen, in {\it Topics in calculus of variations
(Montecatini Terme, 1987)}, Lecture Notes in Mathematics Vol. 1365 (Springer, 
New York,1989).   


\end{thebibliography}
\end{document}